# Constrained scenarios for twenty-first century human population size based on the empirical coupling to economic growth

Barry W. Brook[1,*], Jessie C. Buettel[1] & Sanghyun Hong[1,2]

1. College of Science and Engineering, University of Tasmania, Hobart, Australia
2. *Current Address*: Next Group, 27 Teheran-ro, 2-gil, Gangnam-gu, Seoul, South Korea

\* Corresponding author, email: barry.brook@utas.edu.au

**ORCID**:
Barry W. Brook: 0000-0002-2491-1517
Jessie C. Buettel: 0000-0001-6737-7468
Sanghyun Hong: 0000-0001-6754-2291

**Abstract**
*Growth in the global human population this century will have momentous consequences for societies and the environment. Population growth has come with higher aggregate human welfare, but also climate change and biodiversity loss. Based on the well-established empirical association and plausible causal relationship between economic and population growth, we devised a novel method for forecasting population based on Gross Domestic Product (GDP) per capita. Although not mechanistically causal, our model is intuitive, transparent, replicable, and grounded on historical data. Our central finding is that a richer world is likely to be associated with a lower population, an effect especially pronounced in rapidly developing countries. In our baseline scenario, where GDP per capita follows a business-as-usual trajectory, global population is projected to reach 9.2 billion in 2050 and peak in 2062. With 50% higher annual economic growth, population peaks even earlier, in 2056, and declines to below 8 billion by the end of the century. Without any economic growth after 2020, however, the global population will grow to 9.9 billion in 2050 continue rising thereafter. Economic growth has the largest effect on low-income countries. The gap between the highest and lowest GDP scenarios reaches almost 4 billion by 2100. Education and family planning are important determinants of population growth, but economic growth is also likely to be a driver of slowing population growth by changing incentives for childbearing. Since economic growth could slow population growth, it will offset environmental impacts stemming from higher per-capita consumption of food, water, and energy, and work in tandem with technological innovation.*

## 1. Introduction
Since 1900, the global human population has increased almost five-fold, from 1.7 to 7.7 billion (in 2019) [1,2]. While the rate of growth has been in decline since 1970 [3], over the remainder of the 21st century the world is still likely to gain at least one billion people, and perhaps several billion [4]. The trajectory of the global population this century will have far-reaching implications for social and environmental outcomes [5]. Population growth during the last few centuries has proceeded along with rapid innovation, social advances, and higher aggregate human welfare [6], but also occurring alongside rising carbon dioxide emissions and land-use change, driven by the consumption of food, energy, and other resources [7,8]. Rapid population growth puts pressure on institutions and infrastructure, as well as on natural resources [9,10]. Conversely, stagnating or declining population, which is occurring or predicted to occur in many countries, presents its own challenges [11].

Several studies have forecast the global population through 2050 and 2100, most notably the United Nations Population Division and the World Population Program at the International Institute for Applied Systems Analysis (IIASA) [2,12]. They predict a 2050 population of 9.7 and 9.2 billion, respectively, and a 2100 population of 10.9 and 9 billion. Only IIASA predicts a peak this century, at around 9.4 billion people in around 2070. These forecasts, and indeed most, are based on probabilistic models [13]; that is, their central forecasts are put forward as the most likely population trajectory given a range of uncertainties on the drivers of change, and the uncertainties in their projections can be strongly assumption driven [14].

While such models are useful for long-range planning, population growth ultimately depends on how societies evolve, and there are feedbacks between nations' population and economy, such that a range of possible population futures are plausible [15]. In particular, economic development is strongly related to the so-called demographic transition, wherein countries transition from high fertility and mortality to low fertility and mortality, passing through an intermediate stage of higher fertility than mortality and attendant rapid population growth [16,17]. Consequently, probabilistic modeling can offer only a partial insight into how different pathways of societal change would likely alter the trajectory of the global population.

The connection between economic development and the demographic transition has empirical support in that both fertility and mortality rates are highly correlated with per-capita gross domestic product (GDP), the most robust and commonly used measure of economic development [18,19]. Today, rapid population growth — and the far higher fertility than mortality that it implies — is primarily found in countries with low GDP per capita. Low-income countries as a whole had an average population growth rate of 2.56% in the last five years, compared to 0.47% in high-income countries [2]. Across all countries, GDP per capita is negatively correlated with both fertility and mortality [19,20]. Most countries above a certain GDP-per-capita threshold today have very low mortality rates, due for instance to better access to advanced medical care, but also fertility rates below replacement (of ~2.1 children per woman) [2]. These robust empirical associations make GDP per capita a logical predictor variable for population forecasting.

This economic-demographic transition also has theoretical support from the economic literature, most notably from Unified Growth Theory [18,21,22] [see also 23,24,25], which hypothesizes causal mechanisms. Indeed, the thesis underlying Unified Growth Theory is that economic development — in the form of industrialization and technological progress — changes the underlying motivations and choices around having children (and how many). Most centrally, economic development expands the labor market and raises the relative wage of skilled labor, which incentivizes women to pursue education and formal employment and increases the opportunity cost of childbearing. The incentives of childbearing also change with declining child labor due to shifting social norms and laws as well as declining relative wages of children, acting as an incentive towards improved education [26,27].

Here we exploit the empirical and theoretic association between the demographic transition and economic development (using GDP as a proxy), to create a range of population scenarios under different economic development pathways. Although our method is empirically anchored, employing best-practice statistical cross-validation, model selection, and weighted average forecasting, it is based on a correlative rather than mechanistically causal approach, and is therefore useful for prediction, but not direct attribution or inference [sensu 28]. Nonetheless, our modelling avoids the trap of 'blind' time-series extrapolation [see the admonition by 29], by allowing less developed countries to "learn" from the experience of richer countries, to create plausible forecasts for fertility and mortality that do not simply recapitulate past national trends.

To do this, we first construct a baseline forecast, based on a business-as-usual GDP trajectory [from 30 modelling], and then create a range of new GDP scenarios, modelled at the country level, ranging from no growth (a zero multiplier on annual growth in GDP per capita after 2015), to double annual growth (where the annual GDP per capita growth in the baseline scenario is multiplied by 2) and use these to make corresponding population forecasts. This gives a broad estimate of how population is likely to evolve in response to different socioeconomic pathways. Theory and past empirical results suggest that more rapid growth in per-capita income this century would result in a lower global population, and that slower growth would lead to a higher population. We also test the sensitivity of the global populations forecast to these relationships and trends.

## 2. Methods

We sourced data from the UN Population Division [2] on fertility rates (number of births per woman aged 15 to 44) and mortality rates by age group between 1950 and 2015. Fertility and mortality rates combine to produce what is called natural population growth, which is what we model in this paper; it does not account for immigration or emigration. From James et al. [31] we obtained historical per-capita GDP data, adjusted for purchasing power parity, in constant 2011 international dollars (henceforth termed GDP).

We selected 188 countries (covering 98.9% of the global population in 2015) for which there are comprehensive historical data on fertility and mortality rates and GDP. We classified countries by economic group and geographical region following the World Bank classification [30]. The four economic groups are high, upper middle, lower middle, and low. The seven geographic regions are East Asia and Pacific, Europe and Central Asia, Latin America and the Caribbean, Middle East and North Africa, North America, South Asia, and Sub-Saharan Africa. Such country-, regional- and economic-group-level approaches to forecasting are common in the population-demographic literature [32,33].

A literature review and descriptive statistics were used to formulate possible relationships between GDP and demographic change. Broadly, GDP growth is associated with decreasing fertility and mortality rates, but they decline more slowly at higher levels of GDP. Based on these observations, we used seven monotonic linear regression models including null, linear, negative logarithmic, and negative power functions and three piecewise linear models (linear spline, left-hinge spline, and right-hinge spline models) to encompass a range of plausibly descriptive functional forms that would allow us to predict future fertility and mortality rates in each age group, using the cohort component method, with GDP as an explanatory variable (Table 1), unified as a multi-model average (see below for a full explanation of the weighting scheme for the ensemble forecast).

Our baseline population projection uses the GDP scenario in IIASA's Shared Socioeconomic Pathway 2 (SSP2) where the world's GDP per capita grows from $11,500 in 2015 to $24,715 in 2050 and $41,444 in 2100. SSP2 represents a business-as-usual scenario, where historical trends in economic development continue into the future [34].

To explore the effect of different GDP growth rates on population growth, we generated GDP variations by multiplying the baseline GDP growth in each year and for each country by values between 0 and 2, in increments of 0.1. With the 0 multiplier, GDP levels stay at 2015 levels. With multiplier 1, GDP follows the baseline scenario, SSP2. With multiplier 1.5, global GDP per capita in 2050 is $42,129 and it is $104,201 in 2100; not far from the scenario in SSP5. With multiplier 2, global GDP in 2050 is $74,583 and $284,208 in 2100; this exceeds all SPP scenarios and represents an extreme boundary case rather than a realistic scenario.

**Table 1.** Models used in the population projections, after fitting to historical data on the relationship between population and GPD for each country, where *y* = population size, *x* = GDP, and *μ* = residual error. Model averaging using AIC (information-theoretic) weights was done when using each in the ensemble forecasting.

| Model | Equation | Condition |
|---|---|---|
| Null | $y = \hat{y} + \mu$ | |
| Linear | $y = \beta_1 + \beta_2 \cdot x + \mu$ | |
| Division | $y = \beta_1 + \beta_2 \cdot \frac{1}{x} + \mu$ | |
| Negative logarithmic | $y = \beta_1 + \beta_2 \log_{1/e} x + \mu$ | |
| Negative power | $y = \beta_1 + \beta_2 x^{-e} + \mu$ | |
| Linear spline | $y = \beta_1 + \beta_2 \cdot x_{(x<x_1)} + \mu$ | $(x < x_1)$ |
| | $y = \beta_1 + \beta_2 \cdot x_{(x \geq x_1)} + \mu$ | $(x \geq x_1)$ |
| Right-hinge spline | $y = \beta_1 + \beta_2 \cdot x_{(x<x_1)} + \mu$ | $(x < x_1)$ |
| | $y = \hat{y} + \mu$ | $(x \geq x_1)$ |
| Left-hinge spline | $y = \hat{y} + \mu$ | $(x < x_1)$ |
| | $y = \beta_1 + \beta_2 \cdot x_{(x \geq x_1)} + \mu$ | $(x \geq x_1)$ |

Low-income countries, by definition, have only experienced a narrow range of GDP, and often have not yet experienced a deceleration of demographic change, making inference and forecasting more uncertain based solely on their historical data, and as such likely biasing forecasts over longer time periods. To mitigate this problem, we joined the historical data of a target country with data from countries with higher GDP levels between 1990 and 2015 than a target country's GDP in 2015, but whose maximum GDP between 1990 and 2015 is lower than the maximum GDP pathway of a target country's GDP. This new dataset allowed us to include the experienced relationship between GDP and dependent variables (i.e., fertility and mortality) of economically developed countries in the target country's model while avoiding including the experienced relationship of those countries which have completely different economic conditions.

We then fitted multiple models (Table 1) to these data using maximum likelihood, and calculated the Akaike Information Criterion Weight (*w*AIC$_c$) [35,36] of each model using the target country's historical data. Few countries in our dataset have GDP per capita over $30,000, so model projections much beyond this level involve considerable uncertainty. For this reason, we predicted fertility only up to a GDP level of $30,000, assuming that after this cutoff, fertility stays capped at a constant rate while mortality follows the model prediction. This is consistent with demographic transition theory, where societies' fertility rates reach an equilibrium at high levels of economic development and stop declining thereafter [37]. Setting a threshold is also partly motivated by the observation that beyond around $20,000, most countries show a stable trend in fertility and mortality. We confirmed that setting the threshold at $20,000 or $40,000 did not substantially change the forecasts.

We then multiplied the calculated *w*AIC$_c$ of each model, which is between 0 (no importance) and 1 (most important), with its prediction, to produce an ensemble, information-theoretic-based multi-model forecast, thereby accounting for model-selection uncertainty. The total sum of weighted predictions results in an individual model-averaged prediction for each country, and the aggregate of all countries yields the predicted global human population size, projected from 2015 to 2100.

All modelling was done in R v3.5 [38]. The code and data (.CSV input files) required to replicate all results are available at: https://github.com/BWBrook/pop-gdp

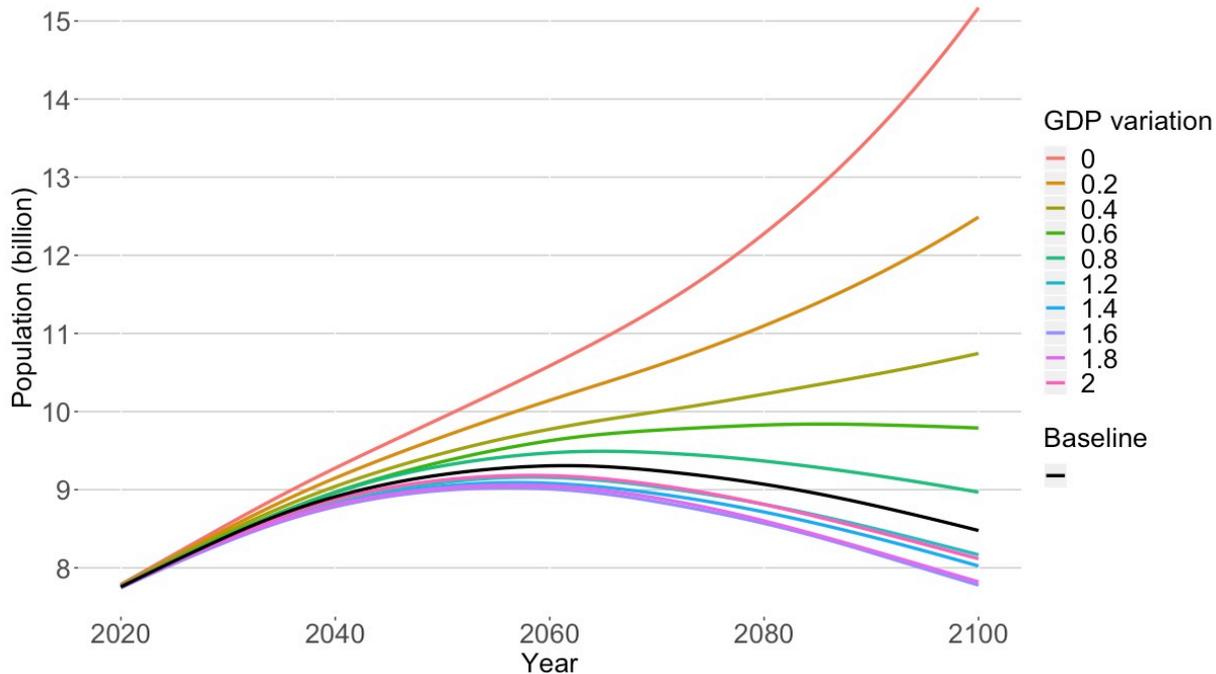

**Figure 1**: Global population projections under different GDP growth (or decline) 'scenarios' for the 21st Century, done for each individual country using a weighted average of the models described in Table 1, and then aggregated across countries for the global count. In terms of GPD variations, 1 is the baseline GDP growth forecast (as predicted by the business-as-usual projection of the Shared Socioeconomic Pathway 2 analysis), and variations represent fractions or multipliers of this baseline. See Methods for details.

## 3. Results
### 3.1 Global
Our baseline forecast predicts that the global population will rise from 7.3 billion in 2015 to 9.2 billion in 2050, peak in 2062 at 9.3 billion and then decline to 8.5 billion by 2100 (Figure 1 and Table 2). However, the rate of growth in GDP per capita has a clear effect on population forecasts. The scenario with the lowest economic growth (with GDP per capita fixed at 2015 levels) sees population projected to grow to 9.9 billion in 2050 and 15 billion by 2100, without any peak this century. Most of the disparity between this forecast and others occurs after 2060. More modest reductions in economic growth from the baseline GDP scenario also result in substantially larger populations over the course of this century. A 50% lower annual GDP growth leads to a population 194 million larger than in the baseline by 2050, a difference that grows to 1.8 billion by 2100.

Conversely, higher-growth scenarios generate far smaller populations. With a GDP multiplier of 1.5 for all countries (meaning 50% higher annual growth in GDP per capita), which results in the lowest population by 2050, the global population barely reaches 9 billion by 2050, and peaks in 2056, declining to 7.8 billion (close to the current [2019] level) by 2100. The difference between the highest and lowest scenario is 940 million by 2050 and 7.4 billion by 2100. Every scenario with a GDP multiplier greater than half of the baseline GDP forecast results in a peaking population this century, and faster GDP growth generally means a lower and sooner peak in population.

Table 2. Population forecasts (millions of people) for three time periods, based on GPD forecasts for each country group, fitted to the models listed in Table 1, and weighted by AIC scores. Results are then aggregated by relative income or geographic region. Also show are the projected peak population and the year reached.

| Area | 2015 | 2050 | 2100 | Peak Pop | Peak Yr |
|---|---|---|---|---|---|
| World | 7278 | 9193 | 8477 | 9307 | 2062 |
| High income | 1158 | 1208 | 1053 | 1223 | 2038 |
| Low income | 630 | 1320 | 2093 | 2093 | 2100 |
| Lower-middle income | 2930 | 4019 | 3660 | 4118 | 2064 |
| Upper-middle income | 2560 | 2646 | 1670 | 2769 | 2034 |
| Latin America and Caribbean | 630 | 760 | 608 | 761 | 2053 |
| South Asia | 1744 | 2355 | 2003 | 2381 | 2060 |
| Sub-Saharan Africa | 955 | 1901 | 2815 | 2815 | 2100 |
| Europe and Central Asia | 905 | 899 | 736 | 919 | 2035 |
| Middle East and North Africa | 429 | 624 | 588 | 640 | 2063 |
| East Asia and Pacific | 2257 | 2263 | 1359 | 2422 | 2031 |
| North America | 357 | 391 | 367 | 391 | 2050 |

In a scenario with global income convergence — where countries below a GDP per capita of $30,000 today (approximately the GDP per capita of Spain) reach this level by 2100, and other countries see no growth — the global population reaches 9.1 billion by 2050, 133 million lower than the baseline. By 2100, the population in this income-convergence scenario reaches 7.7 billion, 785 million less than the baseline and close to the projection in the simple multiplier-1.5 scenario.

While the lowest economic growth scenario consistently results in the largest populations, the converse is not always true: the highest economic growth scenario (multiplier 2) does not always result in the smallest populations. For example, as noted above, the lowest population in 2050 is attained with a global multiplier of 1.5; for 2100, the multiplier with the lowest peak population is 1.9. This stems from the inbuilt assumption in the model that when GDP per capita levels exceed $30,000 a country's fertility rates are held constant, while mortality rates can continue to decline.

*3.2 Income groups*
In the baseline scenario, population will peak and decline this century in all income groups except the low-income group, albeit at different times (Figure 2 and Table 2). The high-income group peaks earliest, followed by the upper-middle and lower-middle income groups, and the low-income group fails to peak at all this century. Irrespective of the specific scenario, most of the growth in the global population this century is projected to be driven by countries currently classified as low income and lower-middle income. Together, these two groups grow by 1.8 billion by 2050 and 2.2 billion by 2100 in the baseline forecast, whereas the upper-middle and high-income groups grow by only 136 million by 2050 and by 2100 have a population almost 1 billion lower than in 2015. As a result, countries currently in the low- and lower-middle-income groups grow from 49% of the global population in 2015 to 58% in 2050 and 68% in 2100.

GDP growth has the largest effect on the low-income group. In these countries, the gap between the highest and lowest GDP growth scenarios is 350 million, more than 25% of the region's population in the baseline scenario (Figure 2). By century's end, the difference enlarges to almost 4 billion. The effect of GDP is still pronounced in the lower-middle income group but to a lesser extent, where the population in low and high GDP scenarios in 2050 differs by over 600 million, or about 15% of the baseline. For the upper-middle and high-income groups, by contrast, the rate of GDP growth has less of an effect, with the difference between high and low GDP growth being only 91 million (3% of baseline forecast) and 40 million (4%) respectively. Since the model assumes fixed fertility after a GDP per capita of $30,000, changes in population in countries above this threshold are driven primarily by ongoing trends in mortality and age structure.

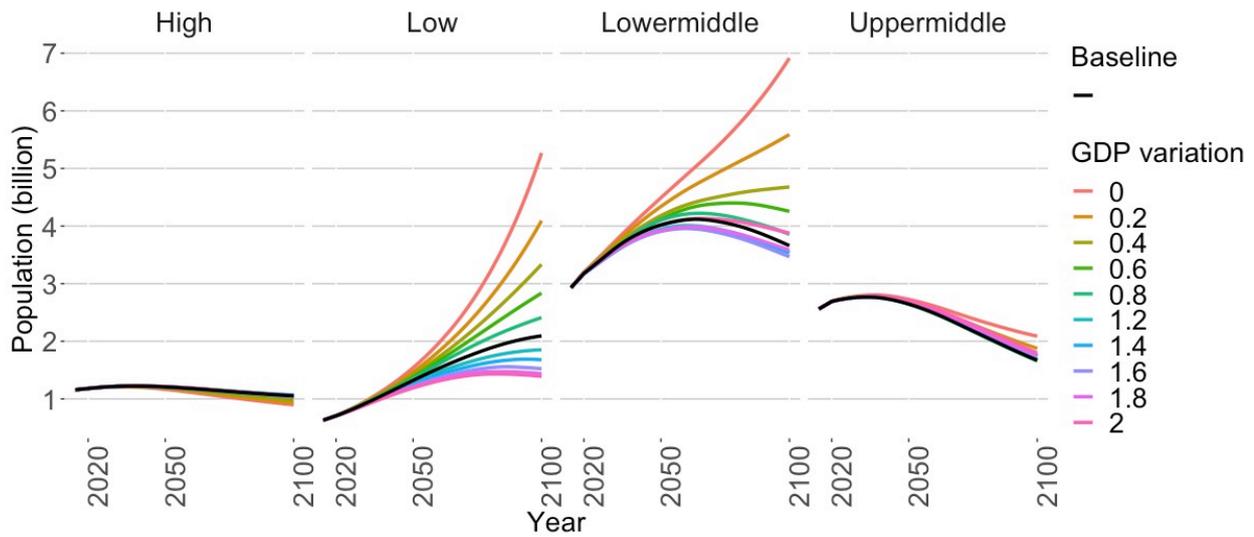

**Figure 2**: Population projections by *income group* under different GDP variations. See Figure 1 caption for further details.

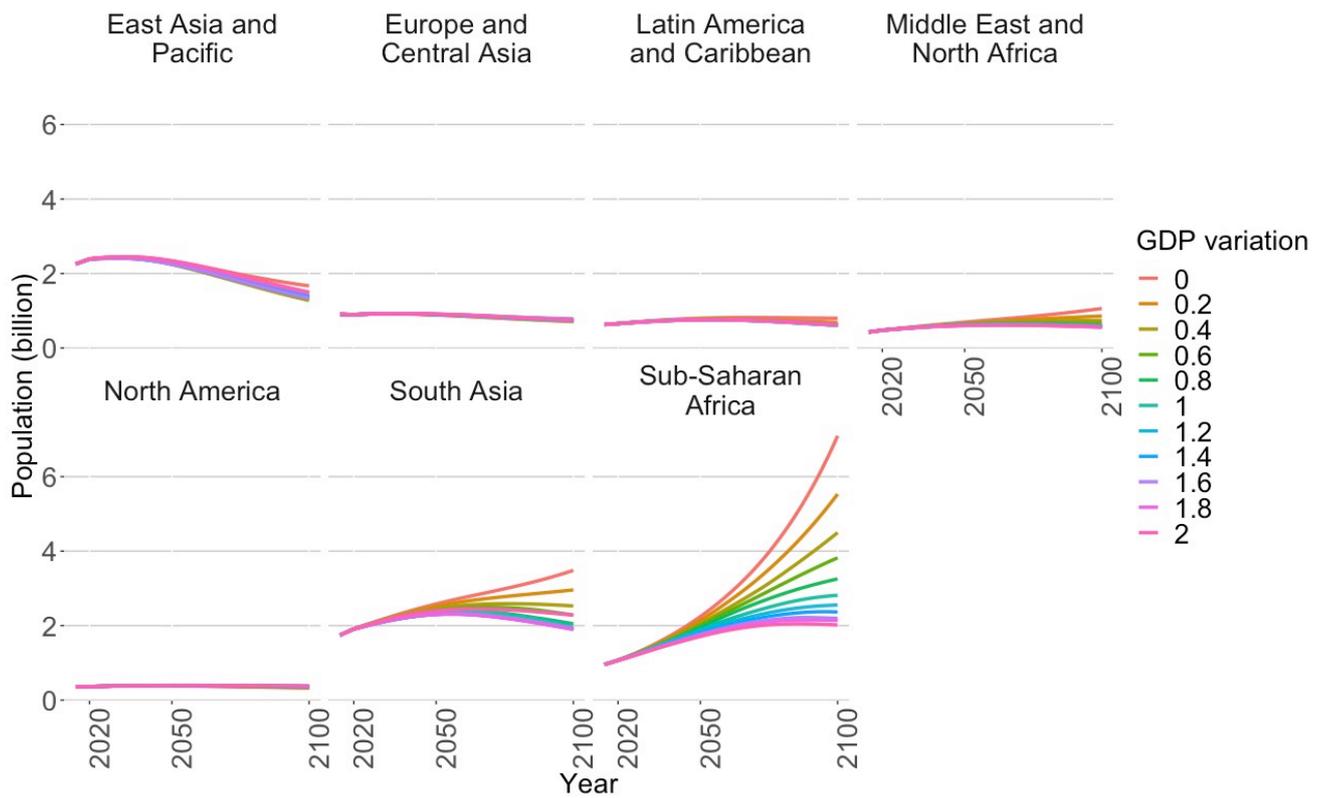

**Figure 3**: Population projections by *geographic region* under different GDP variations. See Figure 1 caption for further details.

*3.3 Regions and countries*

In the baseline scenario, the regions of Sub-Saharan Africa and South Asia experience the largest growth (Figure 3 and Table 2), adding 946 and 611 million people, respectively. While East Asia and the Pacific is currently the region with the largest population, it will be surpassed by South Asia in 2046 and Sub-Saharan Africa in 2059. By 2069, Sub-Saharan Africa is projected to be the region with the largest population, having by then surpassed South Asia. Remarkably, by 2100, Sub-Saharan Africa is expected to make up one-third of the world's population, up from 13% in 2015. The other regions — Europe and Central Asia, Latin America and the Caribbean, Middle East and North Africa, and North America — see more modest growth in absolute terms, with none of these regions growing by more than about 200 million, and all with a peak occurring between 2035 and 2063. Sub-Saharan Africa is the only region whose population is not expected to peak this century.

Sub-Saharan Africa and South Asia also have the largest response to GDP growth (Figure 3). In fact, most of the variation in global population in the global GDP scenarios is driven by these two regions. Without economic growth, South Asia's population would grow to 2.6 billion by 2050, whereas under a high-growth scenario, it would only reach 2.3 billion — a difference of 310 million. For Sub-Saharan Africa, the difference is a far greater 535 million. Of the top 10 countries with the largest difference in population between low and high GDP growth scenarios, 9 are in Sub-Saharan Africa. The sensitivity of population forecasts to GDP growth can be represented as the ratio of the range in population between GDP variation 0 (no economic growth) and 2 (doubled annual economic growth) divided by the baseline forecast for 2050; the map shown in Figure 4 illustrate that the highest ratios for individual countries are concentrated in Sub-Saharan Africa.

Among individual countries, India will undergo the largest growth by 2050 in the baseline scenario (an increase of 366 million relative to 2015), followed by Nigeria (+139 million), Pakistan (+137 million), Ethiopia (+115 million), and Tanzania (+72 million) (Figure 4). The populations of thirty-seven countries are anticipated to shrink by 2050, countries whose current populations account for nearly a third of today's global total, including China (-89 million), Japan (-13 million), Russia (-10 million), Germany (-9.5 million), and Italy (-7.1 million).

**4. Discussion**

We have demonstrated a simple yet robust method for forecasting human population change, that is based on its empirical and theoretical relationships established with GDP per capita. Our method incorporates model-selection uncertainty and weighted forecasting, and it allows less-developed countries to "learn" from the experience of richer countries to ensure plausibility. Although not mechanistically causal, the model is intuitive, transparent, replicable, and anchored to historical data.

The baseline scenario, which follows the GDP growth pattern in the business-as-usual projection of the Shared Socioeconomic Pathway 2, predicts a world population of 9.2 billion by 2050 and 8.5 billion by 2100, with a peak at 9.3 billion in 2062. In this scenario, the populations of all regions but Sub-Saharan Africa and all but the low-income group peak and decline this century. The distribution of population across nations and regions would change radically — in particular, currently

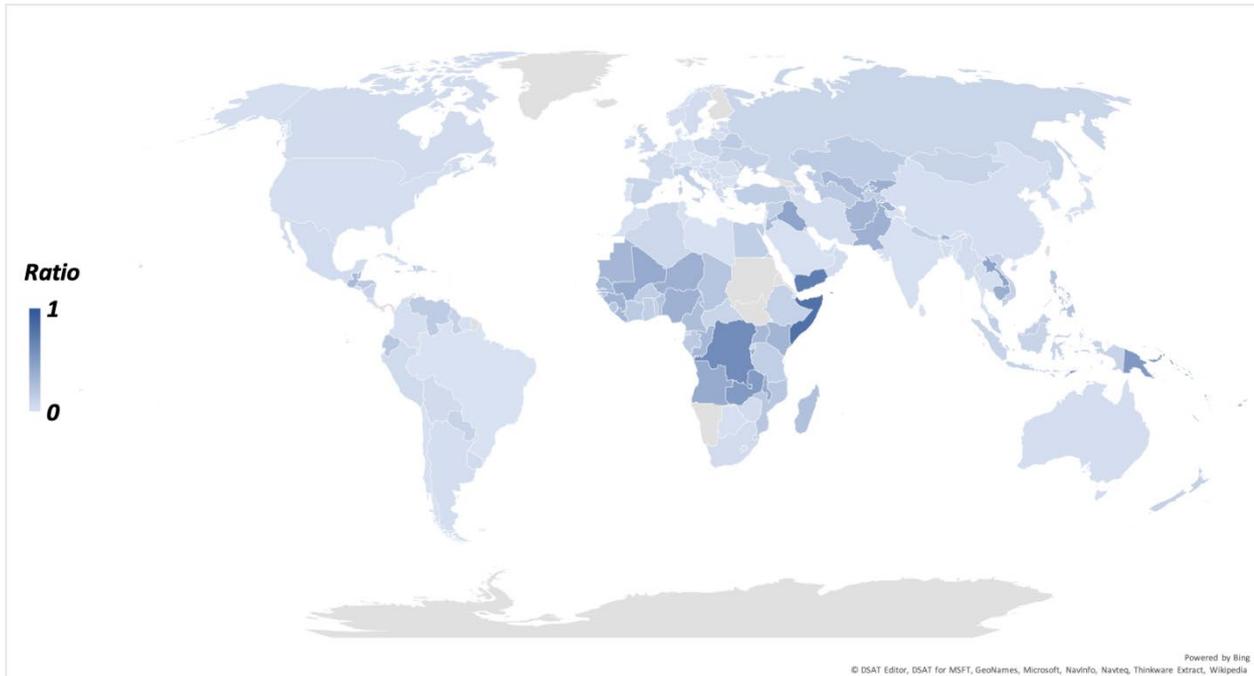

**Figure 4**: Sensitivity of the population forecast to assumptions about future GDP, calculated as the ratio of the range in population between GDP scenario 0 (which assumed stagnation at 2015 levels) and 2 (high growth, double the baseline), divided by the baseline forecast for 2050 (i.e., GPD as predicted by the business-as-usual projection of the Shared Socioeconomic Pathway 2 analysis, see Methods).

low-income countries are going to have a much larger share of the global population.

Notably, the baseline forecast is markedly lower than the mid-range projections of the UN, which predicts 9.7 billion by 2050 and 10.9 by 2100, without a peak this century [1], and are instead more in line with that of IIASA and other earlier calculations [e.g., 29], which predicts 9.2 billion by 2050, a peak at 9.4 billion around 2070, and a 2100 population of 9.0 billion [12].

Given that we based our forecasting method on GPD-growth expectations, the accuracy of our baseline scenario depends on how closely SSP2 approximates future trends in world GDP, as well as the assumption that the historical relationships between population growth and GPD will continue to hold. The fact that our projection under this scenario resembles that of IIASA may in part be an artifact of similar assumptions of future socio-economic development and IIASA's use of projected education, which is correlated with GDP, to predict population growth [39].

Our central finding is that the rate of economic growth will have substantial effects on 21st-century population trajectories. This results from the recurrent historical observation that countries consistently reach a stable equilibrium of low birth and death rates at higher income levels — which is embodied in the model as fixed fertility rates after a GDP per capita of $30,000. Without any economic growth after 2020, population is modelled to grow to 9.9 billion in 2050 and 15 billion in 2100, without a peak. With 50% higher annual growth than the baseline, the world population grows to 9 billion by 2050, peaks in 2056, and reaches, in 2100, a population about the same size as today, at 7.8 billion. We also find that GDP variations have an effect primarily at lower levels of income (Figure 3), which makes Sub-Saharan Africa the region with the most marked sensitivity to GDP growth rates (Figure 4). In short, how much the global population will grow this century depends greatly on realized levels of socioeconomic change, most especially in Sub-Saharan Africa.

Economic development is likely to be

a causal driver of the demographic transition, as outlined in the introduction, via a slew of socio-economic mechanisms and interrelationships. As such, it is likely that policies and emergent societal changes driving economic development will result in lower population growth. However, economic development on its own is not the sole reason that population growth slows. Its partial dependencies, including education and family planning, also act to drive a slowdown in population growth [9,40,41]. While growing economies are better able to fund these services, they can also be expanded independently of economic development. Female empowerment and gender equality have also been identified as important factors at the intersection of demographic and economic change [25,42].

The environmental consequences of different economic futures are ambiguous. Lower population, all else being equal, lowers environmental impacts [10]. However, low population, as our results suggest, is associated with higher per-capita incomes, which drive increased per-capita environmental impacts such as greenhouse-gas emissions [43]. Economic growth is also bi-directionally associated with innovation [44], and could as such lead to development and deployment of less environmentally destructive technologies [45]. Since these trends offset each other, the net effect is not clear.

Our methodology has an arguable advantage over other approaches to population projection in being simple and largely phenomenological and is therefore easily understood and grounded in real-world data. Equally, however, our conclusions come with qualifications. First, caution is merited in interpreting the projections beyond 2050, given that our model parameterization is based on 65 years of data whereas the end year of our forecast is 85 years into the future. Second, while our method for drawing on richer countries' historical experience in creating the models for poorer countries makes our forecasts far more plausible (without it, some countries may see negative fertility rates, for example) compared to time-series models fitted only to historical trajectories, this does involve the assumption that poor countries will follow generally in the footsteps of countries with higher GDP levels. That said, this is the sort of assumption that is virtually unavoidable in any population forecast and is also invoked, within a Bayesian framework, in the UN forecasts [37,46].

It may appear that some of our GDP scenarios are unrealistic, thereby undermining the plausibility of the population trajectories they predict. However, since we set a floor on fertility rates after a GDP per capita of $30,000, even seemingly unlikely GDP trajectories do not necessarily generate implausibly low population projections. In fact, what even the highest GDP per capita scenarios do is simply bring forward the date at which today's poor countries reach a GDP level and demographic transition to a state like that of today's high-income countries. Indeed, as the income-convergence scenario shows, reaching a global human population of less than 8 billion by 2100 requires no economic growth in most developed countries, and only enough growth in poorer nations to bring them to a GDP per capita of about $30,000 by 2100.

Population growth in this century, and the population size at which the human population eventually peaks or stabilizes, will have momentous consequences for societies and the environment. Yet very different population futures are possible [47-49] and, as we have shown here, rates of economic development can lead to vastly different outcomes for humanity this century and beyond, and integrally shape our global environmental footprint.

**Acknowledgments** We thank Linus Blomqvist for providing conceptual input, and he and Kenton de Kirby for critiques on earlier drafts. This work is supported by ARC grant FT160100101 to B.B. **Author Contributions** All authors designed the research and wrote the paper, S.H. did the modelling, S.H. and B.B. analyzed data. **Conflicts of Interest Declaration** None

# References

[1] United Nations Department of Economic Social Affairs - Population Division, The World at Six Billion, 1999.
[2] United Nations Department of Economic Social Affairs - Population Division, World Population Prospects 2019, 2019.
[3] GBD 2017 Population and Fertility Collaborators, The Lancet **392**, 1995 (2018).
[4] United Nations Department of Economic Social Affairs - Population Division, World Population Prospects 2019: Highlights, 2019.
[5] C. Wolf, W. J. Ripple, and E. Crist, Sustainability Science **16**, 1753 (2021).
[6] United Nations Development Program, in *Human Development Reports*2019).
[7] FAOSTAT.
[8] R. K. Pachauri and L. A. Meyer, Climate Change 2014: Synthesis Report. Contribution of Working Groups I, II and III to the Fifth Assessment Report of the Intergovernmental Panel on Climate Change, 2014.
[9] P. Gerland *et al.*, Science **234**, 234 (2014).
[10] C. J. Bradshaw and B. W. Brook, Proceedings of the National Academy of Sciences **111**, 16610 (2014).
[11] D. Bricker and J. Ibbitson, *Empty Planet: The Shock of Global Population Decline* (Robinson, London, 2019), p.^pp. 304.
[12] W. Lutz, W. P. Butz, and S. Kc, *World Population and the Human Capital in the Twenty-First Century* (OUP Oxford, 2014).
[13] A. E. Raftery, N. Li, H. Ševčíková, P. Gerland, and G. K. Heilig, Proceedings of the National Academy of Sciences of the United States of America **109**, 13915 (2012).
[14] D. Rozell, Climatic Change **142**, 521 (2017).
[15] H. Leridon, Population & Societies **573**, 1 (2020).
[16] D. E. Bloom, Science **333**, 562 (2011).
[17] H. Colleran, G. Jasienska, I. Nenko, A. Galbarczyk, and R. Mace, Proceedings of the Royal Society B: Biological Sciences **282** (2015).
[18] O. Galor and D. N. Weil, The American Economic Review **90**, 806 (2000).
[19] D. Herzer, H. Strulik, and S. Vollmer, Journal of Economic Growth **17**, 357 (2012).
[20] L. E. Jones, A. Schoonbroodt, and M. Tertilt, NBER Working Papers, 1 (2008).
[21] M. Cervellati and U. Sunde, SSRN eLibrary, IZA DP No. 2905 (2007).
[22] O. Galor, Cliometrics **6**, 1 (2012).
[23] R. J. Barro and G. S. Becker, in *Econometrica*1989), pp. 481.
[24] G. S. Becker, K. N. Murphy, and R. Tamura, Journal of political economy **98**, S12 (1990).
[25] C. Diebolt and F. Perrin, American Economic Review **103**, 545 (2013).
[26] M. Doepke, Journal of Economic Growth **9**, 347 (2004).
[27] M. Hazan and B. Berdugo, Economic Journal **112**, 810 (2002).
[28] B. Efron, Journal of the American Statistical Association **115**, 636 (2020).
[29] W. Lutz, W. Sanderson, and S. Scherbov, Nature **387**, 803 (1997).
[30] World Bank, 2019).
[31] S. L. James, P. Gubbins, C. J. L. Murray, and E. Gakidou, Population Health Metrics **10**, 12 (2012).
[32] W. Lutz and S. Scherbov, Exploratory extension of IIASA's world population projections: scenarios to 2300, 2008.
[33] S. E. Vollset *et al.*, The Lancet **396**, 1285 (2020).
[34] M. Leimbach, E. Kriegler, N. Roming, and J. Schwanitz, Global Environmental Change **42**, 215 (2017).
[35] K. P. Burnham and D. R. Anderson, Sociological Methods & Research **33**, 261 (2004).
[36] E. J. Wagenmakers and S. Farrell, Psychonomic Bulletin & Review **11**, 192 (2004).
[37] L. Alkema, P. Gerland, A. Raftery, and J. Wilmoth, The United Nations Probabilistic Population Projections: An Introduction to Demographic Forecasting with Uncertainty, 2015.
[38] R Core Team, R: A Language and Environment for Statistical Computing: https://www.R-project.org, 2020.
[39] S. Kc and W. Lutz, Global Environmental Change **42**, 181 (2017).
[40] B. F. Barakat and R. E. Durham, in *World Population and Human Capital in the Twenty-First Century* (Oxford University Press, 2014), pp. 397.
[41] P. J. Gertler and J. W. Molyneaux, Demography **31**, 33 (1994).
[42] H. Kleven and C. Landais, Economica **84**, 180 (2017).
[43] A. Mardani, D. Streimikiene, F. Cavallaro, N. Loganathan, and M. Khoshnoudi, Science of The Total Environment **649**, 31 (2019).
[44] R. P. Maradana, R. P. Pradhan, S. Dash, K. Gaurav, M. Jayakumar, and D. Chatterjee, Journal of Innovation and Entrepreneurship **6**, 1 (2017).
[45] R. York, E. A. Rosa, and T. Dietz, Ecological Economics **46**, 351 (2003).
[46] United Nations Department of Economic Social Affairs - Population Division, World Population Prospects 2017: Methodology of the United Nations Population Estimates and Projections, 2017.
[47] F. Pearce, *The Coming Population Crash: and Our Planet's Surprising Future* (Beacon Press, Boston, 2010).
[48] W. Lutz, W. Sanderson, and S. Scherbov, Nature, 543 (2001).
[49] C. Marchetti, Energy **4**, 1107 (1979).